\documentclass{IEEEtran}
\usepackage{amssymb}
\usepackage{amsfonts}
\usepackage{amsmath}
\usepackage{algpseudocode}
\usepackage{algorithm}
\usepackage{caption}
\usepackage{bm}
\usepackage{graphicx,subfigure,cite,multicol,multirow,diagbox,booktabs,array}
\usepackage{stfloats}

\usepackage{color}
\usepackage{float}
\usepackage{subfig}
\usepackage{makecell}
\usepackage{titlesec}
\ifCLASSINFOpdf
\else
\fi
\begin{document}
	\title{A Novel Tree Model-based DNN to Achieve a High-Resolution DOA Estimation via Massive MIMO receive array}
	\author{Yifan Li, Feng Shu,~\emph{Member},~\emph{IEEE}, Jun Zou, Wei Gao, Yaoliang Song, and Jiangzhou Wang,~\emph{Fellow},~\emph{IEEE}
		\thanks{This work was supported in part by National Key Research and Development Program of China (2022YFE0122300), in part by the National Natural Science Foundation of China (Nos.U22A2002, 62171217 and 62071234), the Hainan Province Science and Technology Special Fund (ZDKJ2021022), the Scientific Research Fund Project of Hainan University under Grant KYQD(ZR)-21008, and the Collaborative Innovation Center of Information Technology, Hainan University (XTCX2022XXC07).}
		\thanks{Y. Li, Jun Zou and Y. Song are with the School of Electronic and Optical Engineering, Nanjing University of Science and Technology, Nanjing 210094, China. (e-mail: liyifan97@foxmail.com).}
		\thanks{F. Shu and Wei Gao are with the School of Information and Communication Engineering, Hainan University, Haikou 570228, China. (e-mail: shufeng0101@163.com).}
		\thanks{J. Wang is with the School of Engineering, University of Kent, Canterbury CT2 7NT, U.K. (e-mail: j.z.wang@kent.ac.uk).}
		\vspace{-2em}}\maketitle
	
	\begin{abstract}
		To satisfy the high-resolution requirements of direction-of-arrival (DOA) estimation, conventional deep neural network (DNN)-based methods using grid idea need to significantly increase the number of output classifications and also produce a huge high model complexity. To address this problem, a multi-level tree-based DNN model (TDNN) is proposed as an alternative , where each level takes small-scale multi-layer neural networks (MLNNs) as nodes to divide the target angular interval into multiple sub-intervals, and each output class is associated to a MLNN at the next level. Then the number of MLNNs is gradually increasing from the first level to the last level, and so increasing the depth of tree will dramatically raise the number of output classes to improve the estimation accuracy. More importantly, this network is extended to make a multi-emitter DOA estimation. Simulation results show that the proposed TDNN performs much better than conventional DNN and root-MUSIC at extremely low signal-to-noise ratio (SNR) with massive MIMO receive array, and can achieve Cramer-Rao lower bound (CRLB). Additionally, in the multi-emitter scenario, the proposed $Q$-TDNN has also made a substantial performance enhancement over DNN and Root-MUSIC, and this gain grows as the number of emitters increases.
	\end{abstract}
	\begin{IEEEkeywords}
		direction-of-arrival (DOA) estimation, deep neural network (DNN), massive MIMO, multi-label learning.\vspace{-0.6em}
	\end{IEEEkeywords}	
	\section{Introduction}
	Direction-of-arrival (DOA) estimation has been a widely studied topic in wireless communications, array signal processing, radar and sonar for few decades \cite{heath2016overview,xu2019novel,chen2021reconfigurable,8663559}. One of the most important directions is how to improve the estimation precision, especially in poor signal conditions such as low signal to noise ratio (SNR) and low number of snapshots. In recent years, with the development of massive MIMO technology, some works considered massive receive arrays for improving the spatial resolution and DOA estimation precision \cite{cheng2015subspace,shu2018low,9570330,li2020covariance,wen2022compressive}.
	
	Nowadays, deep learning (DL) techniques have been introduced into DOA estimation, by matching the principles of DOA estimation with the frameworks of supervised learning. Many high-precision methods can be trained via prior DOA data for various scenarios. \cite{huang2018deep} proposed a deep neural network (DNN) structure for super-resolution channel estimation and DOA estimation based on massive MIMO system.  
	A DNN-based method was also proposed for hybrid massive MIMO systems with uniform circular arrays (UCA) in \cite{hu2019low}. 
	Convolution neural network (CNN) is another technique widely considered in DOA estimation, like \cite{papageorgiou2021deep} designed a CNN-based method for improving precision in low SNR, and \cite{wu2019deep} introduced CNN for DOA estimation with sparse prior. 
	Overall, traditional DL-based DOA estimation methods usually employ a flat single-level network structure \cite{cong2022crb,yu2023deep}, each output class corresponding to a specific direction and the estimation results can only be generated from these fixed directions. So they have high-resolution for on-grid DOAs and have accuracy lower bounds much higher than Cramer-Rao lower bound (CRLB) for off-grid DOAs\cite{wu2022gridless}. 
	
	
	In addition, how to apply the multi-label learning algorithms for the multi-emitter cases is also a key problem in DOA estimation with DL techniques.
	Traditional multi-label learning algorithms like binary relevance, label ranking, multi-class classification can transform complex multi-labeling problems into more accessible single-labeling problems, and methods with another idea are working to apply existing machine learning algorithms to multi-label learning, such as multi label ML-kNN, ML-DT, ML-SVM, etc.\cite{zhang2013review}. Then based on the ideas of multi-labeling learning, \cite{xu2022deep} applied the DNN model for DOA estimation of multi targets, but the DOA estimation accuracy will significantly decrease with the growth of source number.
	
	In order to address the problems of DL-based DOA estimation, a novel DNN architecture called TDNN is studied in this work and the main contributions of this work are summarized as :
    (1) A multi-level tree-based DNN model is proposed, where each level uses small-scale MLNNs as nodes to divide the angular interval into sub-intervals, and each output class is associated to a MLNN at the next level. Then the number of output classes is gradually increasing from the first level to the last level, so TDNN can significantly improve the estimation accuracy by augmenting the depth of the tree instead of enlarging the scale of single network. Simulation results show the proposed TDNN has much better performance than conventional DNN and root-MUSIC at low SNR, and can achieve CRLB as well.  
	(2) TDNN is also extended to multi-emitter scenarios by combining $Q$ parallel TDNNs forms the $Q$-TDNN method. The simulation results show that the proposed $Q$-TDNN has made a substantial performance enhancement over conventional methods.

	\section{System Model}
	Consider $Q$ far-field narrow-band signals impinge onto an $M$-element uniform linear array (ULA). The $q$-th signal is expressed as $s_q(t)e^{j2\pi f_c t}$, where $s_q(t)$ is the baseband signal and $f_c$ is the carrier frequency. Then the output signal at output array is given by
	\begin{equation}
		\mathbf{y}(t)=\sum_{q=1}^{Q}\mathbf{a}(\theta_q)s_q(t)+\mathbf{v}(t)=\mathbf{A}(\boldsymbol{\theta})\mathbf{s}(t)+\mathbf{v}(t),\label{system model}
	\end{equation}
	where $\mathbf{a}(\theta_q)=[1,e^{j\frac{2\pi}{\lambda} d_0\sin \theta_q},\cdots,e^{j\frac{2\pi}{\lambda}(M-1)d_0\sin \theta_q}]^T$ is the array manifold vector, $\mathbf{A}(\boldsymbol{\theta})=[\mathbf{a}(\theta_1),\mathbf{a}(\theta_2),\cdots,\mathbf{a}(\theta_Q)]^T$, $\boldsymbol{\theta}=[\theta_1,\theta_2,\cdots,\theta_Q]^T$ denotes the DOAs to be estimated and $\theta_1<\theta_2<\cdots<\theta_Q$. $\mathbf{v}(t)\sim\mathcal{CN}(\boldsymbol{0},\sigma_v^2\mathbf{I}_M)$ represents the additive white Gaussian noise (AWGN) vector.
	
	Following the signal model (\ref{system model}), the received signal's covariance matrix can be expressed as
	\begin{equation}
		\mathbf{R}=\mathbf{A}\mathbf{R}_s\mathbf{A}^H+\sigma_v^2\mathbf{I}_M=\sum_{q=1}^{Q}\sigma_{s_q}^2\mathbf{a}(\theta_q)\mathbf{a}^H(\theta_q)+\sigma_v^2\mathbf{I}_M,
		\label{covariance matrix}
	\end{equation}
	where $\mathbf{A}=\mathbf{A}(\boldsymbol{\theta})$ and $\mathbf{R}_s={\rm{diag}}\{\sigma_{s_1}^2,\cdots,\sigma_{s_Q}^2\}$ denotes the signal power. However, since $\mathbf{a}(\theta)$ is unknown, the signal covariance matrix cannot be obtained directly, the sample covariance matrix 
	$
	\tilde{\mathbf{R}}=\frac{1}{T}\sum_{t=1}^T \mathbf{y}(t)\mathbf{y}^H(t)
	$
	is considered as a substitution
	and if $T\rightarrow \infty$, we have the sample covariance $\tilde{\mathbf{R}}$ is equal to the statistical one $\mathbf{R}$ in terms of the weak law of large number.
	
	\section{Tree-based DNN model for High-Resolution DOA estimation}
	In this section, a novel tree model-based DNN (TDNN) model is proposed for high-resolution DOA estimation and its detailed training procedure is described as well.\vspace{-1em} 
	\subsection{Proposed Method}
\begin{figure}[t]
		\centering
		\includegraphics[width=0.4\textwidth]{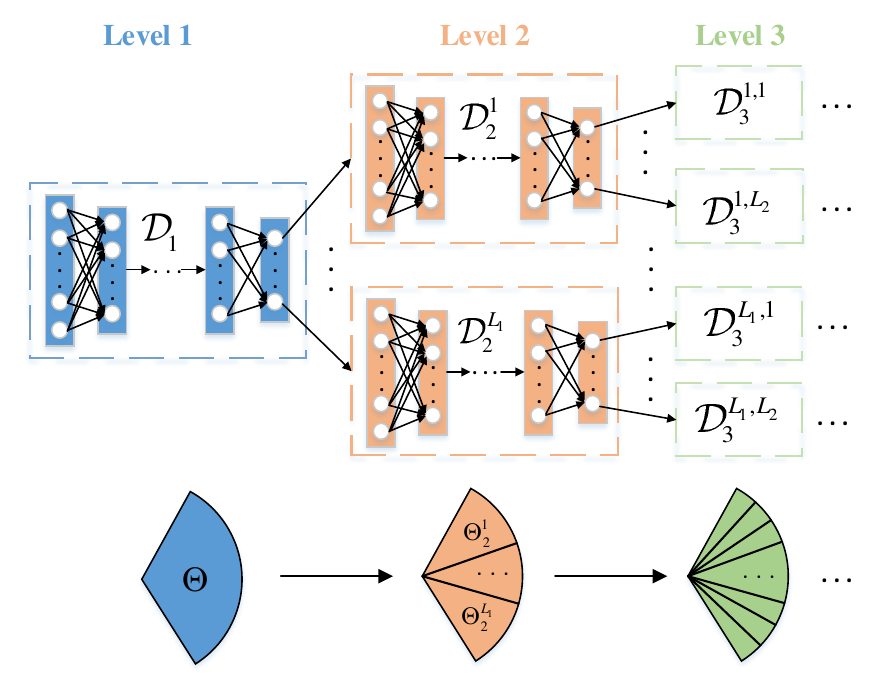}\\
		\caption{Architecture of the proposed TDNN and the process of angular interval divided into sub-intervals by TDNN. \label{ensemble_network}}
		\vspace{-0.10in}	
	\end{figure}

	The proposed TDNN classifier is composed of $H$ levels, and each level contains $G_h$ fully-connected MLNNs, where $1\leq h\leq H$. We let all the $G_h$ networks in the same level of TDNN have identical structures, and their output layers all have $L_h$ neurons. 
	Then the sum output size of level $h$ is $G_hL_h$, which is equal to the number of networks contained in level $(h+1)$, so $G_{h+1}$ can be given by
	\begin{equation}
		G_{h+1}=G_{h}L_{h}=L_1L_2\cdots L_{h-1}L_h,\label{G_h}
	\end{equation}
	for $1\leq h\leq H-1$ and $G_1=1$. 
	As shown in Fig.\ref{ensemble_network}, the feature vector $\mathbf{r}$ is firstly input to $\mathcal{D}_1$, and the output layer of $\mathcal{D}_1$ has $L_1$ neurons. Then in order to join with the output size of level 1, level 2 contains $L_1$ networks, i.e., $G_2=L_1$. So if a signal is divided into the class $l_1$ by $\mathcal{D}_1$, where $1\leq l_1\leq L_1$, its corresponding feature vector $\mathbf{r}$ will be then input to the $l_1$-th network of level 2 ($\mathcal{D}_{2}^{l_1}$). 
	After that, $\mathcal{D}_{2}^{l_1}$ has $L_2$ neurons in its output layer, and there are also $L_2$ networks in level 3 connecting to it. Similarly, if the signal is labelled as class $l_2$ by $\mathcal{D}_{2}^{l_1}$, $1\leq l_2\leq L_2$, $\mathbf{r}$ will be input to $\mathcal{D}_3^{l_1,l_2}$. Therefore, when TDNN is used to perform DOA estimation, only one network in each level will be activated. Thus the input feature vector is sequentially input to $H$ networks, and by combining the classification results of these networks, we can obtain a $H$-dimensional numerical label vector $\boldsymbol{\ell}=[l_1,l_2,\cdots,l_H]^T$, for $1\leq l_h\leq L_h$ and $1\leq h \leq H$. According to $\boldsymbol{\ell}$, the decision procedure that $\mathbf{r}$ experienced in TDNN can be summarized as
	\begin{equation}
		\mathbf{r}\longrightarrow\mathcal{D}_1\stackrel{l_1}\longrightarrow\mathcal{D}_2^{l_1}\stackrel{l_2}\longrightarrow\cdots\stackrel{l_{H-1}}\longrightarrow\mathcal{D}_H^{l_1,l_2,\cdots,l_{H-1}}\stackrel{l_{H}}\longrightarrow \boldsymbol{\ell},
	\end{equation} 
	where the lower right symbol of $\mathcal{D}$ denotes the level number, and the upper right symbols represent the accumulative labels of the previous levels.
	
	Since TDNN is designed for solving DOA estimation problems, we firstly consider the single emitter case, i.e., $Q=1$, and assume the DOA $\theta$ must fall in an angular interval $\Theta=[\theta_{min},\theta_{max})$, i.e. $\theta\in \Theta$, and it is divided into $L_1$ uniform sub-intervals by $\mathcal{D}_1$ like $\Theta=\bigcup_{l_1=1}^{L_1}\Theta_{1}^{l_1}$, where $\Theta_{1}^{l_1}=[\theta_{1,min}^{l_1},\theta_{1,max}^{l_1})$. Then $\Theta_{1}^{l_1}$ can be further divided into smaller sub-intervals as $\Theta_{1}^{l_1}=\bigcup_{l_2=1}^{L_2}\Theta_{2}^{l_1,l_2}$, and $\Theta=\bigcup_{l_1=1}^{L_1}\bigcup_{l_2=1}^{L_2}\Theta_{2}^{l_1,l_2}$, where $\Theta_{2}^{l_1,l_2}=[\theta_{2,min}^{l_1,l_2},\theta_{2,max}^{l_1,l_2})$. Therefore, level $h$ of TDNN can divide $\Theta$ into $L_1L_2\cdots L_h$ uniform spatial subintervals, which is denoted by 
	\begin{equation}
		\Theta=\bigcup_{l_1=1}^{L_1}\bigcup_{l_2=1}^{L_2}\cdots\bigcup_{l_h=1}^{L_h} \Theta_{h}^{l_1,l_2,\cdots,l_h}	
	\end{equation}
	where $\Theta_{h}^{l_1,l_2,\cdots,l_h}=[\theta_{h,min}^{l_1,l_2,\cdots,l_h},\theta_{h,max}^{l_1,l_2,\cdots,l_h})$. And we define the spatial resolution of level $h$ as
	\begin{equation}
		\Delta\theta_h=\theta_{h,max}^{l_1,l_2,\cdots,l_h}-\theta_{h,min}^{l_1,l_2,\cdots,l_h}=\frac{\theta_{max}-\theta_{min}}{G_hL_h},
	\end{equation}
	where $\Delta\theta_1>\Delta\theta_2>\cdots>\Delta\theta_H$.
	Therefore, the size of subintervals divided by the last level, i.e., $\Delta\theta_H=\theta_{H,max}^{l_1,l_2,\cdots,l_H}-\theta_{H,min}^{l_1,l_2,\cdots,l_H}$, represents the spatial resolution can be achieved by TDNN, which is expressed by 
	\begin{equation}
		\Delta \theta=\Delta\theta_H=\frac{\theta_{max}-\theta_{min}}{G_HL_H}=\frac{\theta_{max}-\theta_{min}}{L_1L_2\cdots L_{H-1}L_H},\label{resolution}
	\end{equation}
	and the final estimated DOA $\hat{\theta}$ is related to the classification labels of all the levels, which is given by
	\begin{equation}
		\hat{\theta}=\theta_{min}+\boldsymbol{\ell}^T\Delta\boldsymbol{\theta}=\theta_{min}+\sum_{h=1}^{H}l_h\Delta\theta_h,\label{estimation_result}
	\end{equation}
	where $\Delta\boldsymbol{\theta}=[\Delta\theta_1,\Delta\theta_2,\cdots,\Delta\theta_H]^T$.
	(\ref{resolution}) and (\ref{estimation_result}) show the DOA estimation accuracy of the proposed TDNN classifier depends on the values of $H$ and $L_h$.\vspace{-1em} 
	
	\subsection{Training Procedure}
	All the MLNNs in the proposed TDNN classifier are feed forward networks and composed of three parts, one input layer, one output layer and some hidden layers. Since the networks in the same level has the same number of input and output neurons, in order to reduce the training burden, we let the networks in the same level have the identical network structure and we can train them in the parallel manner. $\mathcal{D}_{h}$ represents an arbitrary network in the level $h$ of TDNN, supposing it has $K$ layers including one input layer, one output layer and $K-2$ hidden layers, then the computation steps of the $k$-th layer in $\mathcal{D}_h$ are given by
	\begin{equation}
		\begin{aligned}
			\mathbf{q}^k_h=\mathbf{W}^{k,k-1}_h\mathbf{g}^{k-1}_h+\mathbf{b}^k_h,~~\mathbf{g}^k_h=A[\mathbf{q}^k_h],
		\end{aligned}
	\end{equation}
	where $1\leq h\leq H$, $1\leq k\leq K$. $\mathbf{g}^k_h$ denotes the output vector in the $k$-th layer of $\mathcal{D}_h$, where $\mathbf{g}^0_h=\mathbf{r}$ and $\mathbf{g}^K_h$ is the output of $\mathcal{D}_h$. $\mathbf{W}^{k,k-1}_h$ represents the fully-connected weight matrix between the $(k-1)$-th layer and $k$-th layer, and $\mathbf{b}^k_h$ is the bias vector. $A[\cdot]$ denotes the activation function, it is set as ReLU function for hidden layers, while Softmax function for output layer.
	
	Referring to \cite{liu2018direction}, the input vector is reformulated by the off-diagonal upper right elements of $\mathbf{R}$, which is given as
	\begin{equation}
		\mathbf{r}=\left[{\rm{Re}}(\bar{\mathbf{r}}^T),{\rm{Im}}(\bar{\mathbf{r}}^T) \right]^T\in \mathbb{R}^{M(M-1)\times 1},
	\end{equation}
	where $\bar{\mathbf{r}}=\left[R_{1,2},\cdots,R_{1,M},R_{2,3},\cdots,R_{2,M},\cdots,R_{M-1,M}\right]^T\in \mathbb{C}^{M(M-1)/2\times 1}$ and $R_{i,j}$ denotes the $(i,j)$-th element of $\mathbf{R}$. In the prediction period, since the covariance matrix is unavailable, the testset consists of the off-diagonal upper right elements of the sample covariance matrix $\tilde{\mathbf{R}}$.

	Since MLNNs in level $h$ have same structures, let $\mathbf{z}^h$ represent the one-hot form label vector for training $\mathcal{D}_h$, which is an $L_h\times 1$ binary vector with $\lVert\mathbf{z}^h\rVert=1$ and its relationship with $\boldsymbol{\ell}$ can be expressed as 
	$\mathbf{z}^h(\boldsymbol{\ell}(h))=\mathbf{z}^h(l_h)=1$.
	Therefore, by combining the training data and training label, the complete training set for the $i$-th MLNN is given as $\mathbb{T}=\{(\mathbf{r},\mathbf{z}^{h})\}$. Then we define $\hat{\mathbf{z}}^{h}$ as the output prediction vector of $\mathcal{D}_h$ for $\mathbf{r}$, which is in the form of probability distribution. So we choose binary cross entropy (BCE) as the loss function and the optimal weights and biases of $\mathcal{D}_h$ can be obtained by minimizing
	\begin{equation}
		\begin{aligned}
			loss=-\frac{1}{L_h}&\sum_{l_h=1}^{L_h}\bigg[\mathbf{z}^{h}(l_h)\log(\hat{\mathbf{z}}^{h}(l_h))+\\
			&(1-\mathbf{z}^{h}(l_h))\log(1-\hat{\mathbf{z}}^{h}(l_h))\bigg].
		\end{aligned}
	\end{equation}
\vspace{-1em} 
	\subsection{Complexity Analysis}
	In the prediction stage, only one MLNN in each level of TDNN is activated for per DOA estimation, then the complexity of TDNN comes from two parts, i.e., model complexity and computation complexity. Firstly, we define the model complexity as the number of output classes required for achieving a specific resolution $\Delta\theta$, so the model complexity of conventional DNN is
	$\mathcal{M}({\rm{DNN}})=O\left(N\right)$,
	where $N=(\theta_{max}-\theta_{min})/\Delta\theta$. Then the model complexity of TDNN is given by
	\begin{equation}
		\mathcal{M}({\rm{TDNN}})=O\left(\sum_{h=1}^H L_h\right),
	\end{equation} 
	from (\ref{resolution}) we know $N=L_1L_2\cdots L_H$, so it is obvious that $\mathcal{M}({\rm{TDNN}})\ll \mathcal{M}({\rm{DNN}})$.
	
	As discussed in \cite{hu2019low}, the computation complexity of a network is proportional to the number of layers and the number of neurons in each layer. So as the summation of all the activated MLNNs, the computation complexity of TDNN is 
	\begin{equation}
		\mathcal{C}({\rm{TDNN}})=O\left(\sum_{h=1}^{H}\sum_{k=1}^{K-1}W_h^{k}W_h^{k+1}\right),\label{time complexity}
	\end{equation}
	where $W_h^{k}$ denotes the number of neurons contained in the $K$-th layer of $\mathcal{D}_h$. 
	
	\begin{figure}[ht]
		\centering
		\includegraphics[width=0.33\textwidth]{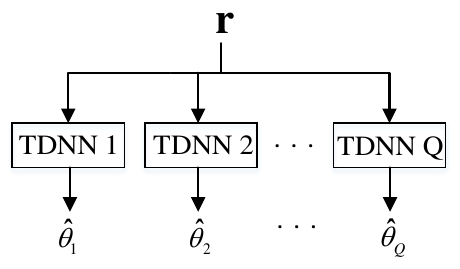}\\
		\caption{Proposed $Q$-TDNN for multi-emitter DOA estimation.\label{Q-INN}}
		\vspace{-0.20in}
	\end{figure}
	\section{Extensive TDNN-based Method for Multi-Emitter Scenarios}
	As the number of signal sources increases, the corresponding DOA estimation problem is converted into a multi-label learning problem. 
	For conventional DNN-based DOA estimation algorithms, each unit in the output layer corresponds to an independent angle, so when the combined signals from $Q$ different sources are input to the DNN, the training label vector will be a binary vector $\mathbf{z}_{\rm{DNN}}$ which contains $Q$ '1' elements, i.e., $\lVert\mathbf{z}_{\rm{DNN}}\rVert_1=Q$. 
	However, for our proposed TDNN, since the output neurons of each level correspond to different angular regions, it is possible that several DOAs in the same region under the multi-emitter case, then there will appear that different signals marked by same label and thus the binary vector cannot express it. Therefore, we give an $L_h\times Q$ binary label matrix for classifying the $Q$-emitters signals as
	$\mathbf{Z}^h=\left[\mathbf{z}_1^h,\mathbf{z}_2^h,\cdots,\mathbf{z}_Q^h\right]$,
	where $\mathbf{z}_q^h$ denotes the one-hot form label vector of the $q$-th signal and $\lVert\mathbf{z}_{q}^h\rVert_1=1$.
	
	Since the feature vector $\mathbf{r}$ is only constructed by the upper right elements of $\mathbf{R}$ and has no connection with noise, then by observing (\ref{covariance matrix}) that $\mathbf{r}$ can be separated into $Q$ components under $Q$-sources cases as
	$\mathbf{r}=\mathbf{r}_1+\mathbf{r}_2+\cdots+\mathbf{r}_Q$,
	where $\mathbf{r}_q$ is the feature component of $\theta_q$ and its elements are from the upper right part of $\sigma_{s_q}^2\mathbf{a}(\theta_q)\mathbf{a}^H(\theta_q)$.
	Therefore, by drawing on the idea of binary relevance, we consider transforming the $Q$-emitters DOA estimation problem into $Q$ single-source problems. And when solving the $q$-th problem, we can regard $\mathbf{r}_q$ as the principal component and $\mathbf{r}$ is classified as $\theta_q$. Then the $Q$-TDNN algorithm is proposed based on this principle.
	As shown in Fig.\ref{Q-INN}, $Q$-TDNN is composed of $Q$ TDNNs with same structures, and TDNN $q$ is used to solve the $q$-th problem. Firstly, $\mathbf{r}$ is separately input to these $Q$ TDNNs, $\boldsymbol{\ell}_q=[l_{q,1},l_{q,2},\cdots,l_{q,H}]^T$ denotes classification result of TDNN $q$, then based on (\ref{estimation_result}) we can obtain the estimation result of $\theta_q$ as 
	\begin{equation}
		\hat{\theta}_q=\theta_{min}+\boldsymbol{\ell}_q^T\Delta\boldsymbol{\theta}=\theta_{min}+\sum_{h=1}^{H}l_{q,h}\Delta\theta_h.
	\end{equation}
	
		
		\begin{table}[htb]
			\centering 
			\caption{Model parameters.}\label{parameters}
			\setlength{\tabcolsep}{2pt}
			\begin{tabular}{c|c|c|ccc}
				\toprule [1pt]    
				\multirow{1}{*}{} & \multirow{1}{*}{$H$} & \multirow{1}{*}{$G_h$} & input layer & hidden layers & output layer \\ 
				\hline  
				\multirow{2}*{2-Level TDNN} & \multirow{2}*{2} & 1 & \multirow{2}*{4032} & \multirow{2}*{512,256,128,64,32,16}&12\\~ & ~ &12 &~& ~&10\\\hline 
				\multirow{3}*{3-Level TDNN} & \multirow{3}*{3} & 1 & \multirow{3}*{4032} & \multirow{3}*{512,256,128,64,32,16}&6\\~ & ~ &6 &~& ~&5\\~ & ~ &30 &~& ~&4\\
				\bottomrule [1pt]     
			\end{tabular}			
		\end{table}
	
	\begin{figure*}[tbp]
		\centering
		\begin{minipage}{0.32\linewidth}
			\centering
			\includegraphics[width=0.9\linewidth]{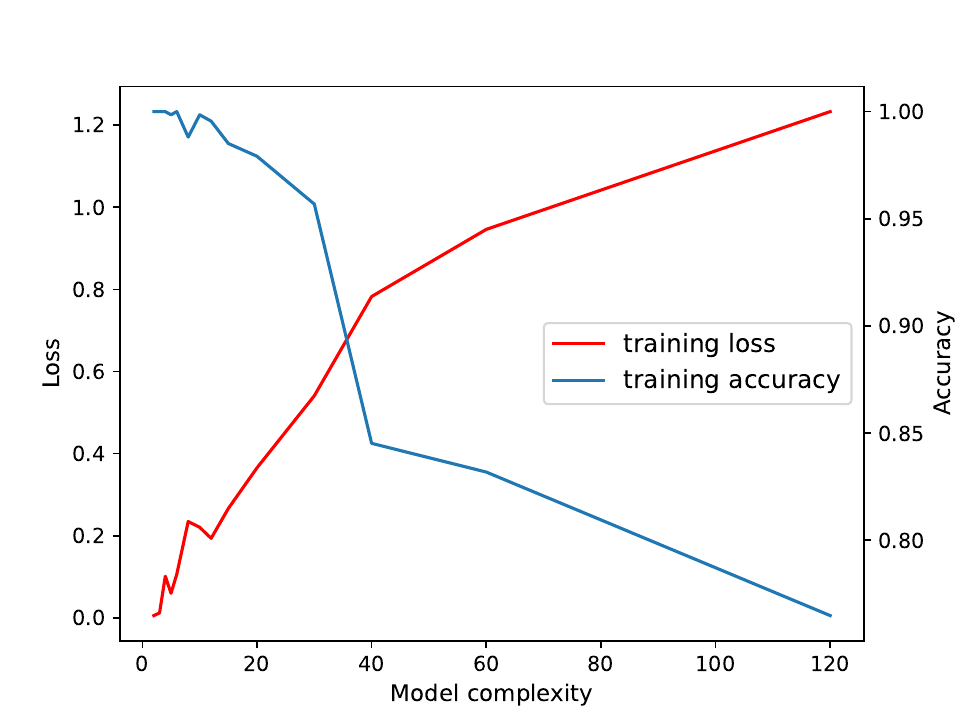}
			\caption{Training performance versus model complexity}
			\label{training performance}
		\end{minipage}
		\begin{minipage}{0.32\linewidth}
			\centering
			\includegraphics[width=0.9\linewidth]{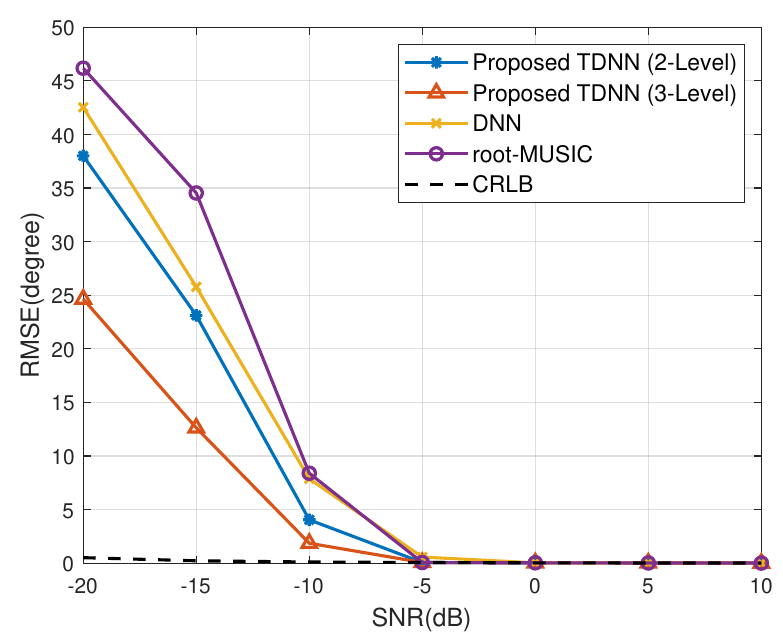}
			\caption{RMSE versus SNR.}\label{rmse_snr}
		\end{minipage}
	\begin{minipage}{0.32\linewidth}
		\centering
		\includegraphics[width=0.9\linewidth]{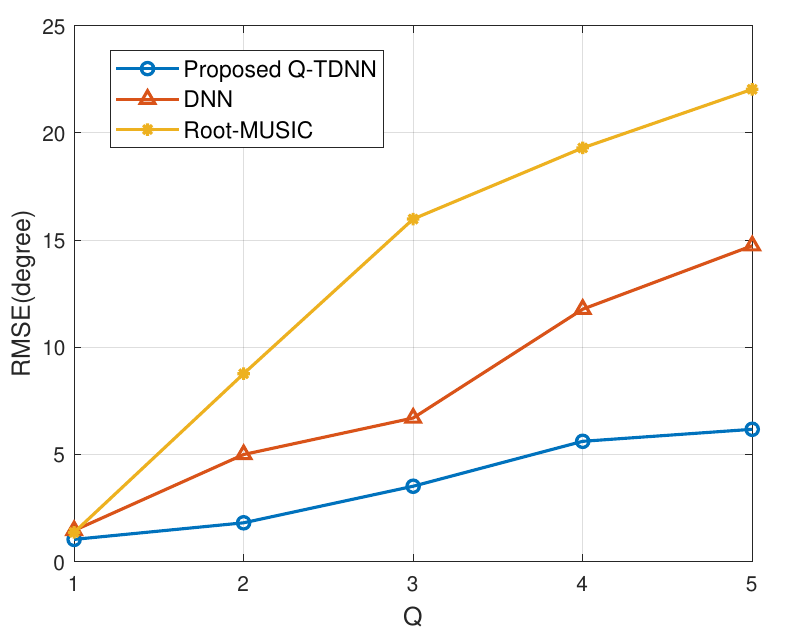}\\
		\caption{RMSE versus Q.}\label{rmse_Q}	
	\end{minipage}
	\end{figure*}
	
		\section{Simulation Results}
		In this section, the DOA estimation performance of the proposed methods is verified by performing a series of numerical simulations. The receive array is equipped with a 64-elements ULA, the interelement spacing of the ULA is
		half-wavelength, and the DOAs of all the signals are fall in the angular region $\Theta=[-60^{\circ},60^{\circ})$. 
		These simulations are implemented on Tensorflow. The proposed TDNN method is mainly compared to DNN \cite{xu2022deep} and root-MUSIC\cite{barabell1983improving}.
		
		Table \ref{parameters} lists the model parameters of the proposed TDNN, where $H$ denotes the depth of TDNN and $G_h$ represents the number of MLNNs contained in each level. The network structures of MLNNs in the 2-level TDNN and 3-level TDNN are also shown in this Table. 
		
		Fig.\ref{training performance} plots the training performance versus the model complexity of DNNs. From this figure, it is seen that the model accuracy  of DNNs becomes worse with the increasing of output classes. This means small-scale networks can have better model accuracy than large-scale networks. The specific impact on DOA estimation performance will be analyzed in the following by combining Fig.\ref{rmse_snr}. 
		
		Fig.\ref{rmse_snr} demonstrates the DOA RMSE versus SNR of the proposed TDNN with DNN, root-MUSIC and CRLB. The specific parameters involved in this simulation are as follows: $N=50$, $Q=1$ and $\theta=27^{\circ}$. From this figure, it can be seen  that TDNN attains obvious performance gains over conventional DNN and root-MUSIC at $\rm{SNR}\leq -5dB$. TDNN can also nearly achieve CRLB at $\rm{SNR}=-5dB$, while DNN can only achieve it at $\rm{SNR}=0 dB$. Additionally, the comparison between TDNNs with different structures is noteworthy. The performance of 3-level TDNN is better than 2-level TDNN, and from Table \ref{parameters} we can find the model complexity of 3-level TDNN is also lower than 2-level TDNN. Therefore, by combining the conclusion of Fig.\ref{training performance}, we can conclude lower model complexity makes higher DOA estimation accuracy, and it is a key index must be considered when a TDNN model is constructed.
		
		Fig.\ref{rmse_Q} depicts the DOA estimation RMSE versus of $Q$-TDNN, DNN and root-MUSIC under the multi-emitter scenarios and $\rm{SNR}=-8dB$. It can be seen from this figure the RMSE of these three methods are very close when $Q=1$, but as the number of emitters increases, the performance gap between the other two methods and TDNN is increasing, when $Q=5$,  $Q$-TDNN has about $16^{\circ}$ performance advantage over root-MUSIC and $8.5^{\circ}$ over DNN. Therefore, the proposed $Q$-TDNN is confirmed to be a much more accurate method for the multi-emitter cases.
		
		
		\section{Conclusions}
		In this work, a novel multi-level tree-based DNN model for high-resolution DOA estimation with massive MIMO receive array was proposed. Each level of TDNN adopts small-scale MLNNs as nodes to partition the target angular interval into multiple sub-intervals and each output class is associated to a MLNN at the next level. As the number of MLNNs increases from the first level to the last level, and so as the sum output classes. Therefore, TDNN can improve the DOA estimation accuracy by adding the number of levels without increasing the model complexity of each MLNN. The proposed TDNN performs much better than conventional methods like Root-MUSIC, DNN when SNR is in the extremely low region ($<$-5dB). Additionally, $Q$-TDNN method is also designed for multi-emitter scenarios on basis of TDNN. The performance enhancement of $Q$-TDNN over DNN and root-MUSIC also increases as the number of emitters increases.
		

		\bibliographystyle{IEEEtran}
		\bibliography{INN}
	\end{document}